\documentclass{ws-procs9x6}

\newcommand{\Heebar}{$^4\overline{\rm He}$}
\newcommand{\Hee}{$^4$He}
\newcommand{\He} {$^3$He}

\newcommand{\pbar}{$\overline{\rm p}$} 
\newcommand{\tbar}{$^3\overline{\rm H}$} 
\newcommand{\dbar}{$\overline{\rm d}$}
\newcommand{\hypertbar}{$^3_{\bar{\Lambda}} \overline{\rm H}$}
\newcommand{\Hebar}{$^3\overline{\rm He}$}
\newcommand{\hypert}{$^3_{\Lambda}H$}
\newcommand{\Libar}{$^6\overline{\rm Li}$}
\newcommand{\Lbar}{$\overline{\Lambda}$}

\begin{document}

\title{Phenomenological study of light (anti)nuclei, (anti)hypertriton and di-Lambda production at RHIC}

\author{L. Xue, Y. G. Ma$^*$,  J. H. Chen, S. Zhang}
\address{Shanghai Institute of Applied Physics, Chinese Academy of Sciences, Shanghai 201800, China\\
$^*$E-mail: ygma@sinap.ac.cn}

\begin{abstract}
We present the production of light (anti)nuclei, (anti)hypertriton and di-Lambda based on coalescence model in central Au+Au collisions at $\sqrt{s_{NN}}=200GeV$. The invariant yields of \He(\Hebar), \hypert(\hypertbar), \Hee(\Heebar) obtained is found to be consistent with the STAR measurements \cite{STARproton,STARhyperons,STARdHe3,H3Lbar,He4bar}. The $p_{T}$ integrated yields for di-Lambda $dN_{\Lambda\Lambda}/dy \sim 2.23\times10^{-5}$, and is not strongly dependent on the parameter employed for coalescence process. Relative particle ratios of light anti(nuclei) and (anti)hypertriton are explored, and agree with experimental data and thermal model predictions \cite{H3Lbar,He4bar,Phenixdbar,STARpbarpRatio} quite well. An exponential reducion behavior is presented for the differential invariant yields with increased baryon number. The production rate reduces by a factor of 1692 (1285) for each additional antinucleon (nucleon) added to antinuclei (nuclei), and the production rate of \Libar~ is predicted to be $10^{-16}$ which is consistent with STAR result \cite{He4bar}.

\end{abstract}


\bodymatter

\section{Introduction}
Antimatter light nuclei \dbar~, \tbar~, \Hebar~have been widely studied in both cosmic rays and  nuclear reactions for the purposes of dark matter exploration and the study of manmade matter such as quark gluon plasma (QGP). On the other hand, searching for antihypernuclei and di-baryon bound states and exploring the Y-N and Y-Y interactions have been steadily fascinated the sights of physicists. In this paper, we report an investigation of the production of light (anti)nuclei and (anti)hypertriton as well as di-Lambda in relativistic heavy-ion collisions base on the coalescence model \cite{Sato}.

\section{Production of light (anti)nuclei and (anti)hypertriton and di-Lambda}
\begin{figure}[htb]
\centering
\makebox[0cm]{\includegraphics[width=0.8 \textwidth]{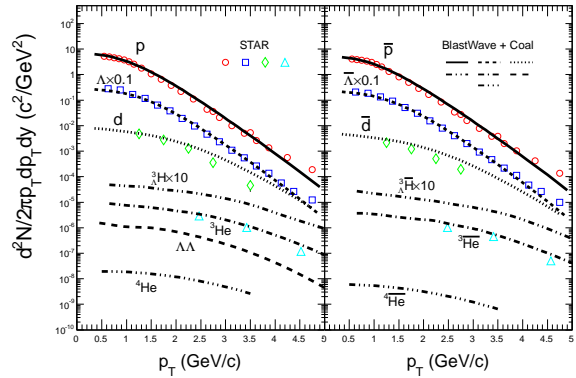}}
\caption{ Differential invariant yields versus $p_{T}$ distributions for p(\pbar), $\Lambda$(\Lbar), light (anti)nuclei, as well as (anti)hypertriton and di-Lambda. The open symbols are experimental data points from the STAR measurement\cite{STARproton,STARhyperons,STARdHe3}, and the black lines represent the preliminary results from hydrodynamical BlastWave model plus a coalescence model.}
\label{lightnuclei}
\end{figure}

We focus on central Au+Au collisions at RHIC energy in our work. Figure \ref{lightnuclei} shows the calculated differential yields of p(\pbar), $\Lambda$(\Lbar), light (anti)nuclei as well as (anti)hypertriton and di-Lambda versus transverse momentum ($p_{T}$) distribution. Our calculations reproduce the data extracted by the STAR experiment \cite{STARproton,STARhyperons,STARdHe3}, and make predictions for the production rates of \hypert~(\hypertbar), \Hee~(\Heebar) and di-Lambda. The $p_{T}$ integrated yields (dN/dy) for \hypertbar, \Heebar, and di-Lambda are about $4.9\times10^{-5}$, $1.10\times10^{-7}$, and $2.23\times10^{-5}$ respectively. The production of $\Lambda\Lambda$ is not strongly dependent on the parameters employed for coalescence model.

\begin{figure}[h]    
\centering
\makebox[0cm]{\includegraphics[width=0.7 \textwidth]{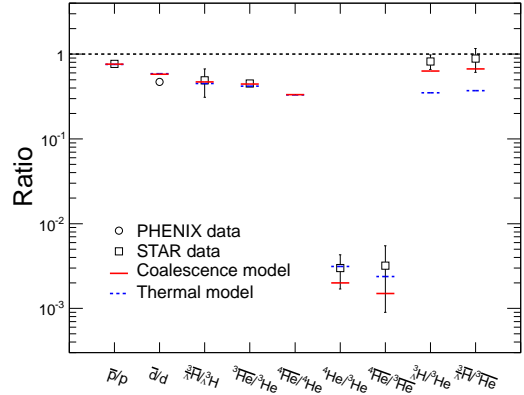}}
\caption{ Comparison of particle ratios between data and model calculations. Open symbols represent the data from STAR and PHENIX experiments \cite{H3Lbar,He4bar,Phenixdbar,STARpbarpRatio}. The solid lines and dashed lines are the preliminary results from our calculation and thermal model prediction\cite{PBM} respectively.}
\label{ParticleRatios}
\end{figure}

Relative particle ratios of light (anti)nuclei and (anti)hypernuclei are studied and compared with the RHIC data and thermal model predictions as depicted in Figure \ref{ParticleRatios}. Our results can fit the antinuclei to nuclei ratios as well as \Hee/\He~and \Heebar/\Hebar~ at RHIC energy quite well, and are also consistent with thermal results. For \hypert/\He~and \hypertbar/\Hebar~, the coalescence model has a better description than thermal model.

\begin{figure}[tp]    
\centering
\makebox[0cm]{\includegraphics[width=0.7 \textwidth]{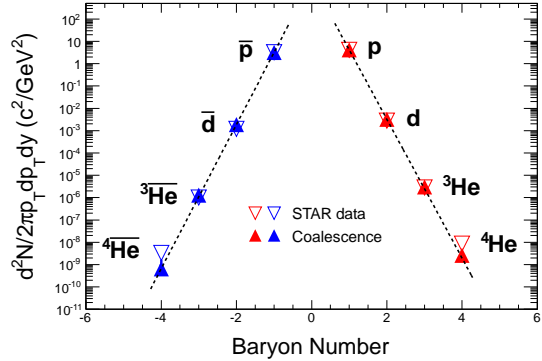}}
\caption{ Invariant yields $d^{2}N/(2\pi p_{T}dp_{T}dy)$ of (anti)nucleus in the average transverse momentum region ($p_{T}/|B| = 0.875GeV/c$) as a function of baryon number (B). The open symbols represent the data points extracted by the STAR experiment at RHIC energy, while solid ones are the preliminary results reproduced by coalescence model.  The lines represent the exponential fit for the coalescence results of positive particles (right) and negative particles (left) with formula $e^{-r|B|}$.}
\label{InvariantYieldvsB}
\end{figure}

Figure~\ref{InvariantYieldvsB} presents a decreasing exponential trend of differential invariant yields with increased baryon number. The reduction factor obtained by fitting the distribution is a number of 1692 (1285)  for each additional antinucleon (nucleon) added, and is comparable with STAR data\cite{He4bar}. By extrapolating the distribution to B = -6 region, we predict that the production rate of next stable antimatter nuclei \Libar~ is about $10^{-16}$. On the other hand, the excitation of light (anti)nuclei from a highly correlated vacuum is discussed somewhere else \cite{newphysics}. This new production mechanism can be investigated by comparing their invariant yields from models and experimental data. Any deviation of the experimental data point from the model expectation should be a hint of the new production mechanism. Our results are consistent with the STAR measurement within uncertainties, and do not support the hypothesis of the excitation production from the vacuum.

\section{Summary}
In summary, we presented a vigorous calculation for the production of light (anti)nuclei, (anti)hypertriton and di-Lambda, based on the coalescence model in central Au+Au collisions at $\sqrt{s_{NN}}=200GeV$.  We demonstrate that the current approach can reproduce the differential invariant yields and relative production abundances of light (anti)nuclei and (anti)hypernuclei. The $p_{T}$ integrated yields (dN/dy) for di-Lambda is about $2.23\times10^{-5}$.
The exponential behavior of the differential invariant yields versus baryon number distribution is studied. By extrapolating the distribution to B = -6 region, the production rate of \Libar~ in high energy heavy ion collisions is about $10^{-16}$¡£

\section{Acknowledgments}
This work is partially supported by the NSFC under contracts No.
11035009, No. 10979074, No. 10905085, and No. 11105207,  the Knowledge Innovation Project of Chinese Academy of Sciences under Grant No. KJCX2-EW-N01, and Dr. J. H. Chen is partially supported by the Shanghai Rising Star Project under Grand No. 11QA1408000.


\bibliographystyle{ws-procs9x6}
\bibliography{ws-pro-sample}

\end{document}